\documentclass[print]{ati}
\usepackage{graphicx,times}
\usepackage{lmodern}
\usepackage[sort&compress, numbers,super]{natbib} 
\bibpunct[,]{[}{]}{,}{n}{}{,}			
\usepackage{url}
\usepackage[colorlinks=true,breaklinks=true, linkcolor=red,urlcolor=magenta,citecolor=blue,anchorcolor=green]{hyperref}
\usepackage{caption}
\usepackage{rotating}
\usepackage{CJKutf8}	
\usepackage{fancyhdr}
\renewcommand{\citet}[1]{\citeauthor{#1}(\citeyear{#1})\cite{#1}}	

\usepackage{amsmath}
\usepackage{color}

\usepackage{hyperref}
\usepackage{listings}

\begin{document}

   \title{The Operation Control System for the Tianlai Experiment}

   \author{Jixia Li
      \inst{1,2}
      \ORCID{0000-0001-9652-1377}
   \and Fengquan Wu\correspondingAuthor{}
     \inst{1,2}
     \ORCID{0000-0002-6174-8640}
   \and Shijie Sun
      \inst{1,2}
      \ORCID{0000-0003-2775-3523}
   \and Yougang Wang
      \inst{1,2}
      \ORCID{0000-0003-0631-568X}
   \and Xuelei Chen\correspondingAuthor{}
      \inst{1,2}
      \ORCID{0000-0001-6475-8863}
   }
\correspondentEmail{wufq@bao.ac.cn, xuelei@cosmology.bao.ac.cn}

\institute{
    State Key Laboratory of Radio Astronomy and Technology, National Astronomical Observatories, CAS, A20 Datun Road, Chaoyang District, Beijing, 100101, P. R. China;
    \and
    School of Astronomy and Space Science, University of Chinese Academy of Sciences, Beijing 100049, China;
   }
   \date{Received:~ 2025;   Accepted:~2025;  Published Online:~2025; 
   \DOI{ati} }			
   \citeinfo {Li, J. X. et al.}\volume{0}\issue{0} \pages{0--1}	
   \StartPage{0} 			
   \MonthIssue{0}		
   \copyrights {2025}     		
   \abstract{
The Tianlai 21cm intensity mapping experiment is located at the Hongliuxia Observing Station, which is a remote site with excellent electromagnetic environment. To facilitate the operation of the Tianlai experiment while reducing the required human power and travel cost, we have designed the system to be remotely controllable from the start. 
In this paper, we present the basic design of the operation control system, including the control network, and the controlling mechanisms for the power switch, the steering of the dish antenna, the analog and digital components of the array, and the operation of the array. In the design of this system of operation control, we emphasize the following points: online accessible, simplicity, flexibility, strict control of electromagnetic interference (EMI) and security. The various devices are connected in a local area network (LAN), and one can control them remotely by securely logging into a server on the LAN and issue commands. We describe the functions of the programs designed for the control. Similar design and the various hardware and software components may also be applicable or of reference value to other remote observing stations. 
  }

   \authorrunning{ASTRONOMICAL TECHNIQUES \& INSTRUMENTS }   
   \titlerunning{Li J. X. et al.: Tianlai Observation System}  
   \maketitle
   \setcounter{page}{\Page}	
%
%

\section{Introduction}
\label{sec:intro}

The Hongliuxia Observing Station, located in the Balikun county, Hami, Xinjing Uygur Autonomous Region, is the site for the Tianlai experiment, which aims to develop and test key techniques for the 21cm intensity mapping observation of large scale neutral hydrogen distribution  \citep{Chenxl2012tianlai,xuyd2015forecast}. As the cosmic 21cm signal from the neutral hydrogen is extremely weak, and the red-shifted 21cm line lays in the low frequency part of the radio spectrum, where the radio frequency interference (RFI) is usually quite severe, it is vital for the site to have minimal RFI, so as to allow the best measurement possible. The Hongliuxia station is a site with excellent electromagnetic environment, and good logistic support conditions \citep{Wufq2014site}, but it is relatively remote. 

To facilitate the regular operation of the telescope instruments, while reducing the number of people needed to man the station and run the experiment, it is essential to design a set of internet-based remote monitoring and control system for this observing station. Other observatories may also encounter similar problems.

As the astronomical observatories or telescopes have many different variations and functions, their control systems are also vastly different \citep{control-overview,fast-central-control}. While there are a number of general-purpose telescope control software and standards—such as the Instrument Neutral Distributed Interface (INDI\footnote{ https://indilib.org}), Robotic Telescope System 2 (RTS2\footnote{ https://rts2.org}), Astronomy Common Object Model (ASCOM\footnote{ https://ascom-standards.org}), Observatory Control and Astronomical Analysis Software (OCAAS) \citep{OCAAS}, and Automatic Telescope Instruction Set (ATIS) \citep{ATIS}, in many cases these do not address the unique requirements of specific telescope systems, and specialized design are made. For example, in some telescopes, high precision is required, thus the precision control is the focus in the design of the control system \citep{kongdeqing-pointing, ZiWu2}, which may need to take into account of mechanical deformation, temperature deformation  \citep{kunming40m-control,QTT110m}, and active surface control  \citep{tianma-distribution-control,tianma-control,fast-progress,fast-active-control}. In other cases,  the convenient switching of the multiple feeds, receivers and digital backends \citep{kunming40m-feed-switch,NanShan-feed-switch,GBT}, or the synchronization of the many antennas in an array may be the major concern  \citep{mingantu-control,mingantu-muser,fast-realtime,alma-realtime,ska-control,alma-control}.

While many astronomical observatories are built in remote, sparsely populated areas for their better observing conditions, up to now most still require a regular staff for operation and maintenance, the fully remote-control approach as we adopted is still an unusual one. For example, we consider the following observing stations in China: 21CMA, MUSER, and AIMS.

The 21 Centimeter Array (21CMA).
Located in Ulastai, Xinjiang, is a radio telescope array dedicated to epoch of reionization 21cm observations \citep{21CMA_HY,21CMA_ZhQ}. It was constructed in 2004, and its antennas and receivers are fixed with no active control needs. A simple remote monitoring system for the backend servers was developed, but without full remote control capability for the entire observation system. Recently, with the newly added pulsars backend, a remote operation and management system is under construction\footnote{Private communication by the 21CMA team}.

The Mingantu Spectral Radioheliograph (MUSER). 
A solar radio telescope array located in Zhengxiangbaiqi, Inner Mongolia, is another typical remote observing station \citep{MUSER}. Its routine operation relies on a permanent rotating on-site duty team\footnote{Private communication by the MUSER team}.

The Accurate Infrared Magnetic field Measurements of the Sun (AIMS). This is a newly built optical/infrared telescope in Lenghu, Qinghai, with a network-based modular control architecture  \citep{AIMS_DengYY}. However, the infrared imaging system require regular liquid nitrogen refilling, so a regular operation and maintenance staff is required for daily operations\footnote{Private communication by the AIMS team}, and RFI is not an issue that require attention for this telescope.

In this paper, we present the design and implementation of the Tianlai experiment control system, which is an integration of a hybrid set of hardware and software. It provides a simple, efficient solution to our main problem, namely the operation of an observatory at a remote place with low cost. 
It allows us to control the station remotely and automatically, with minimal amount of human intervention, and maintains the excellent RFI environment of the site. 

The construction of the Tianlai experiments, including the station house, the pathfinder antenna arrays, and associated facilities such as the power line and optical fiber, as well as the installation of the electronics and other instruments have been completed by the end of 2016. It has been running since then as a pathfinder experiment, and many of the controlling devices and software are added or upgraded in the mean time according to our need.

The performance of the experiments including the stability, were reported in \cite{Lijx2020tianlai,Wu2021dish}. 
Various works have been carried out with the facility, including hardware design and development \cite{liut2014feed,chenzhp2016cyldesign,alecksander2017feed,niuchh2019roach,Wangzh2024roach}, algorithm, software and data analysis \cite{zhangj2016sky,zuoshf2019eigen,Zuosf2021tlpipe,anh2022algoscr,olivier2022dish,yukf2023simu,yukf2024skymap,santanu2018progress,daic2019rfi}, drone measurement of beam \cite{Zhangjy2021dishbeam,lijx2025drone}, cross-coupling effect \cite{lijx2021reflect,sunshj2022em,juhun2024coup,liuyf2025beam}, beam-forming and fast radio burst (FRB) detection \cite{yuzj2022dishfrb,YuZJ2024cylfrb}.
The experiment has been designed as a pathfinder experiment for SKA. We are also expanding the experiment to include 21cm global spectrum measurements \cite{zhu202121cm,Sunshj2024cali,Xujq2025GSM,zhang2025precision}, test of membrane antenna\cite{Suonanben2024}, and outrigger antennas to allow better determination of FRB locations. 
The remote control system has played an important part in these works, and will be naturally extended to cover these updates.

Based on the situation of the facilities at Hongliuxia Observing Station, we have the following considerations in designing the remote control system. 

\begin{itemize}
\item{\bf Online Accessible.}
As the Hongliuxia station is located at a remote place, from the beginning we designed the system to allow most operational commands be carried out remotely, using internet connection.
    
\item{\bf Simplicity.}
The operations of the Tianlai arrays are relatively simple, with only a few modes, and the observations are mostly drift scan. We have therefore chosen to design a simple system, with the most essential and frequently used commands in an uniform interface, which is easy to learn and use for the operation of the array. We also favor the use of commonly available hardware components.

\item{\bf Flexibility.}
Many different kinds of hardware from various manufactures are used at the site, to perform a large number of functions. The design needs to be flexible to accommodate these heterogeneous devices, and easy to adjust with addition of new components. 

\item{\bf Strict Control of EMI.}
The electromagnetic interference (EMI) generated by the hardware can affect the detection of the weak signal we are searching, so we make every effort to mitigate this source of RFI. We have separated the station house and antenna sites, and the analog and digit department rooms. We also designed the ancillary devices at the antenna site such as the antenna drive motors to be powered off during the drift scan observations. 

\item{\bf Security.}
The system needs to be securely protected, to ensure the safe operation of the instruments.

\end{itemize}

This paper is structured as follows. In Sec. \ref{sec:facility}, we briefly introduce the Tianlai site and instruments. In Sec. \ref{sec:network}, we describe our control network. In Sec. \ref{sec:power}, we describe the power control, including the general method, the UPS system, and the power of the antenna site. In Sec. \ref{sec:ant_ctrl}, we discuss the antenna steering control. In Sec. \ref{sec:array}, we present the various system and software for the operational control of the array. Finally, we conclude in Sec. \ref{sec:conclusion}

\section{The Tianlai Facilities}
\label{sec:facility}

\begin{figure*}[ht]
    \centering
    \includegraphics[width=0.99\linewidth]{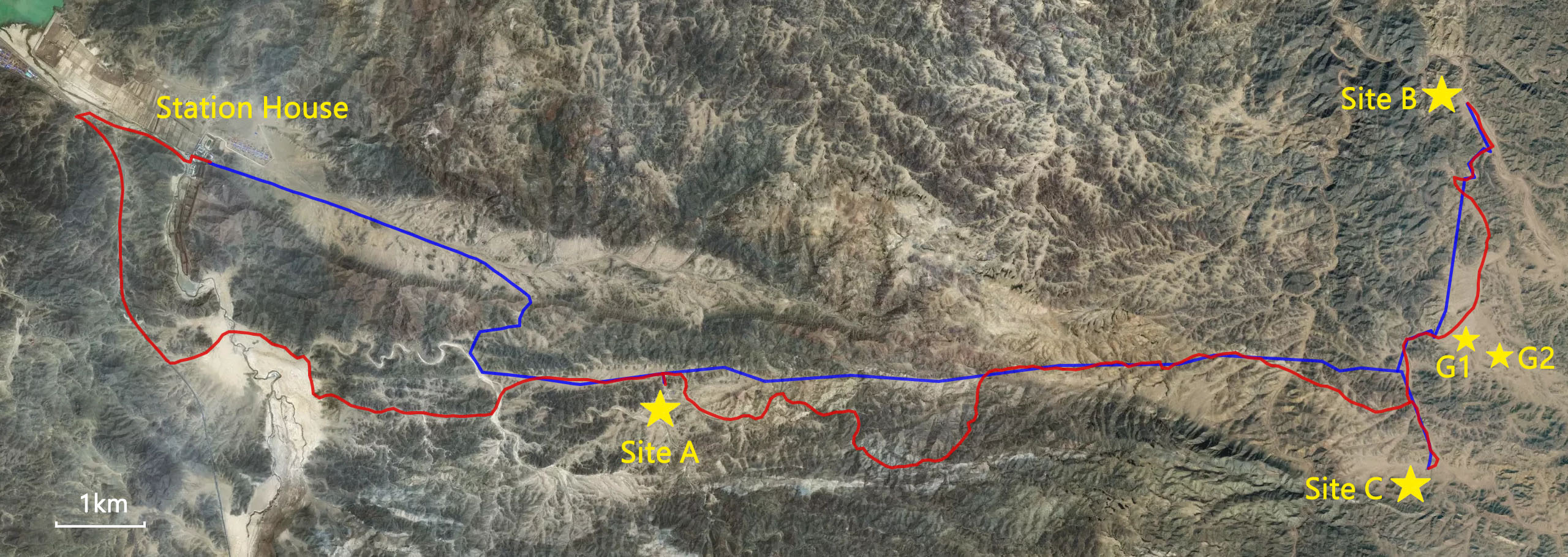}
    \caption{Location of the station house, the existing antenna site (Site A), future FRB outriggers (Site B, C), and global spectrum experiments (Site G1, G2). The blue curve shows the optic fiber route and the red curve shows the road to each site.  }
    \label{fig:site_map}
\end{figure*}

The Hongliuxia Observing Station comprises a station house located in the nearby Da Hongliuxia Village, along with the Cylinder Array and the Dish Array. The antenna arrays are located at a site about 6 km away from the station house as crow flies. This antenna site is connected to the station house by a 10 kV power line, as well as an optical cable with multiple fibers. It can be accessed from the village by a dirt road about 11 km long. The map of the station is shown in Fig. \ref{fig:site_map}, where Site A is the antenna site. The blue line indicates the optic cable and power line, and the red line indicates the dirt road. Recently, we have also set up new sites further east, for installing global spectrum experiments (marked as G1, G2 on the map), and FRB detection outriggers (marked B, C on the map).

Site A is surrounded by low hills. The relatively long distance between the antenna sites and the village ensures that it is minimally affected by the RFI generated in the village. The station house provides the living quarters for researchers, and also have rooms for most of the analog and digital electronics of the arrays. It is located in the village, which is more convenient (e.g. allow easy access to water supply) and safer during the winter when the dirt road is covered by heavy snow and becomes difficult to access. Internet connection service is provided at the village. 

\subsection{The Antenna Site}

The antenna site (currently site A) includes the cylinder array, dish array, and a calibration noise source (CNS). These are powered by a linear power transformer. 

The cylinder array consists of three adjacent parabolic cylinder reflectors, each 40 m $\times$ 15 m, with their long axes oriented in the N-S direction. Dual-polarization dipole feeds are placed along the focus line of each cylinder. The cylinders are closely spaced in the east-west (E-W) direction. The reflectors focus the incoming radio signal in the E-W direction, while allowing a wide field of view (FoV) in the N-S direction. 

The operation of the cylinder array is relatively simple. The cylinders are fixed on the ground. They passively collect the signal, and as the Earth rotates, the beams make drift scan survey of the northern celestial hemisphere. 

Very recently, three smaller cylinders of 24 m $\times$ 15 m have been constructed. One is located at site A, and one each at site B and site C, though their installation is not yet complete. These will work as FRB outriggers and help improve FRB localization precision.  

The dish array originally consists of 16 dish antennas of 6 m aperture, each equipped with a dual linear-polarization receiver feed at the primary focus, and arranged in two concentric rings with one antenna at the center. Most recently, two more 6m dishes are added at site A but about 100 meters away from the original dishes, to provide longer baselines for calibration. The dishes are mounted on Alt-Azimuth mounts, and can be steered to any direction in the sky above the horizon by its motor drive. 

The operation of the dish array is also relatively simple, in the normal observation mode, we just point the dishes to a fixed direction (usually along the meridian) and make drift scan observations. We usually do not carry out tracking observations, as these dish drivers are not designed for tracking with high precision. 
However, for calibration, we also steer the antenna to some celestial targets, then observe them to transit over the beam.

\begin{figure*}
    \centering
    \includegraphics[width=0.99\linewidth]{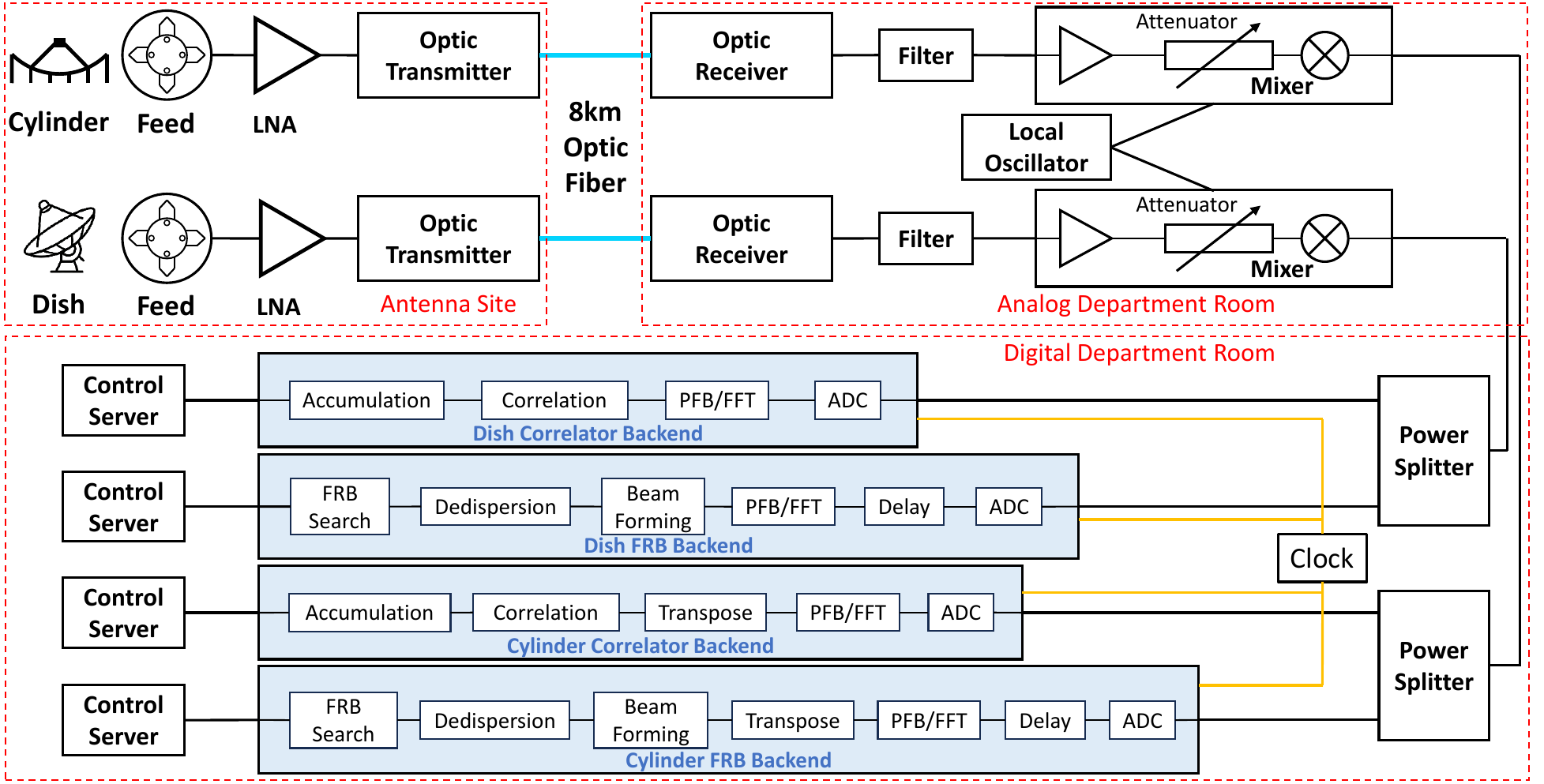}
    \caption{The various parts and signal flow of the Tianlai experiment.
    }
    \label{fig:signal_schematic}
\end{figure*}

In Fig. \ref{fig:signal_schematic}, we show a schematic of the signal flow of the arrays.
The sky signal collected by the antenna are first amplified by the low noise amplifier (LNA) on the feed. Then the radio frequency (RF) analog signal is converted to optical signal at the antenna site, and then transmitted by the optic fiber to the station house. 

At the site, there is also CNS, which is an omnidirectional antenna located in a nearby hill at site A. It can be turned on, to broadcast a noise-like signal in the observing band periodically, which can be used to make relative calibration of the instrument phase during the drift scan observations \cite{zuoshf2019eigen}. However, we found from the observational data that during night time, even without this calibration the instrument phases are fairly stable.

\subsection{The Station House}

The Station House includes the living quarters and the analog department and digital department. The living quarters include dormitories, conference rooms, kitchen, bathrooms, storage areas, and a garage. The living quarter is mostly controlled by the people on site directly, though the electric heating system for the house can also be remotely controlled. The remote control is mainly applied to the devices in the instrument rooms. 

\subsubsection{Analog Department}

The analog department is located in the analog department room which houses the optic receivers, the local oscillator (LO), mixers, the radio frequency (RF) and intermediate frequency (IF) amplifiers, and filters, attenuators, and related components, as shown in Fig.~\ref{fig:signal_schematic}. Optic cables from the antennas are drawn into the analog department room, where the signal is firstly transferred back to the electric signal, and then fed into analog signal chain. This room is especially sensitive to RFI. Separation of the analog and digital department rooms help to prevent the RFI generated by the digital devices from entering the analog part of the circuit.  

Each optic receiver has 36 channels. A total of 7 optic receivers are installed inside the analog department room, one for the dish array and six for the cylinder array, each can handle 36 input signals.  The optic receivers are designed to start working automatically once powered on. They convert the optical signal to electric RF signal which are fed to the mixers. Each mixer is also injected with an oscillation RF signal from the split output of a single local oscillator (LO), to down convert the frequency to intermediate frequency (IF) range. The mixers are grouped as units of  12 channels. For the dish antennas 3 mixers are used, and for the cylinder antennas 17 mixers are used (one of them is a redundancy mixer for malfunctioning channels). The level of the RF, LO and IF signal can be adjusted by attenuators for equalization. The attenuator units are controlled electronically with a DB9 socket, with the commands defined by the manufacture, and can be communicated using the RS-232 protocol. During the commissioning time, or later when necessary, we adjust the attenuation according to the measurements with external instruments or the readings of correlators. All optic receivers and the mixers are installed in three 19-inch standard cabinets. 

The IF signals are transmitted out to the digital department room through well sealed RF connectors installed in the wall between the two rooms.

\subsubsection{Digit Department}

The digit department is located in the digital department room which includes the correlators, servers, network switches, and related equipment. It currently contains four main digital systems: the dish correlator backend, the dish FRB backend, the cylinder correlator backend, and the cylinder FRB backend. Each IF signal from the analog department room is firstly split into two paths via a power splitter. One path is sent to the correlator backend for interferometry, while the other is sent to the corresponding FRB backend for transient detection. This setup enables both scientific operations, 
interferometry and FRB detection, to be conducted simultaneously. 

The dish correlator backend consists of 3 FPGA boards. They are enclosed in one chassis. Two FPGA boards are used for AD, FFT, correlation and integration. One FPGA board is used for the CNS control, clock synchronization and system trigger. The firmware have been burnt into the chips, so once powered on, the correlator will boot into ready status and waiting for observation parameters. One control server is used to send out the parameters, and receives and saves the visibility data into HDF5 files. The format of the files are detailed in  \citet{Zuosf2021tlpipe}. 

The dish FRB backend is made up by a B-engine (beam-former) and D-engine (de-dispersion server) \citep{yuzj2022dishfrb}. The B-engine has two CASPER SNAP2 \footnote{\href{https://casper.astro.berkeley.edu/wiki/Hardware}{https://casper.astro.berkeley.edu/wiki/Hardware}} chassis, which convert the analog signal to digital (AD), Fast Fourier Transform (FFT) and the digital beam-forming. The D-engine includes two GPU servers to conduct de-dispersion and FRB searching. In addition, a control server is used for managing the dish FRB detector system.

The cylinder correlator backend is made up by the F-engine and X-engine \citep{Wangzh2024roach}. The F-engine consists of 6 ROACH-2 servers, which digitize the signal and make the FFT, then the data are transmitted to the X-engine via a 64-port 10~Gb switch. The X-engine consists of 7 GPU servers, each of which will correlate the data of the same frequency and integrating into visibilities. Finally, the visibilities from the 7 GPU servers of different frequency channels are collected and re-assembled in the control server, and the data are dumped into hard disks in HDF5 format \citep{Zuosf2021tlpipe}. 

The cylinder FRB backend consists of an F-engine which performs FFT to convert the data from the time domain to the frequency domain, and a B-engine which forms beams at each frequency by a weighted summation of the signals from different feeds, and a D-engine which searches the de-dispersed data stream for FRB candidates \citep{YuZJ2024cylfrb}. The F-engine has 24 FPGA boards, each is equipped with 4 ADCs and a Xilinx Kintex-7 FPGA. The B-engine contains 12 GPU servers, each has 2 GPU nodes inside. The data between the F-engine and the B-engine are exchanged by two 48-port 10~Gb switches. The beam-forming data are transmitted to the D-engine using a 48-port 10~Gb switch. Finally, the D-engine de-disperses the beam data and searches for FRB candidates using 6 GPU servers. Two control servers have been setup for system boot up and monitoring. One is used for the F-engine and B-engine, and the other one is used for the D-engine.

Besides the above observing systems, there is also a Network Time Protocol (NTP) time server in the room to provide 10 MHz reference clock for the 4 sets of digital systems.

\subsubsection{Auxiliary Department}

The auxiliary department includes devices which provide infrastructure for the steady operation of all the facilities, like the normal temperature adjusted by the air conditioner, the stable power adjusted by the UPS.

Precision air conditioners are installed in the analog department room (1 unit) and digit department room (2 units). Each air conditioner (A/C) unit is remotely accessible through a DB9 socket using the RS-485 protocol, which enables us to monitor the temperature and humidity of the room. 

An Uninterruptible Power Supply (UPS) unit is installed in the digit department room. It provides a stable AC power for the devices in both the digit department room and the analog department room. Its control is described in more detail in Sec.\ref{sec:power}. 

In addition, some surveillance cameras are installed to monitor the operation of the two department rooms.

\section{The Control Network}
\label{sec:network}

\begin{figure*}
    \centering
    \includegraphics[width=\linewidth]{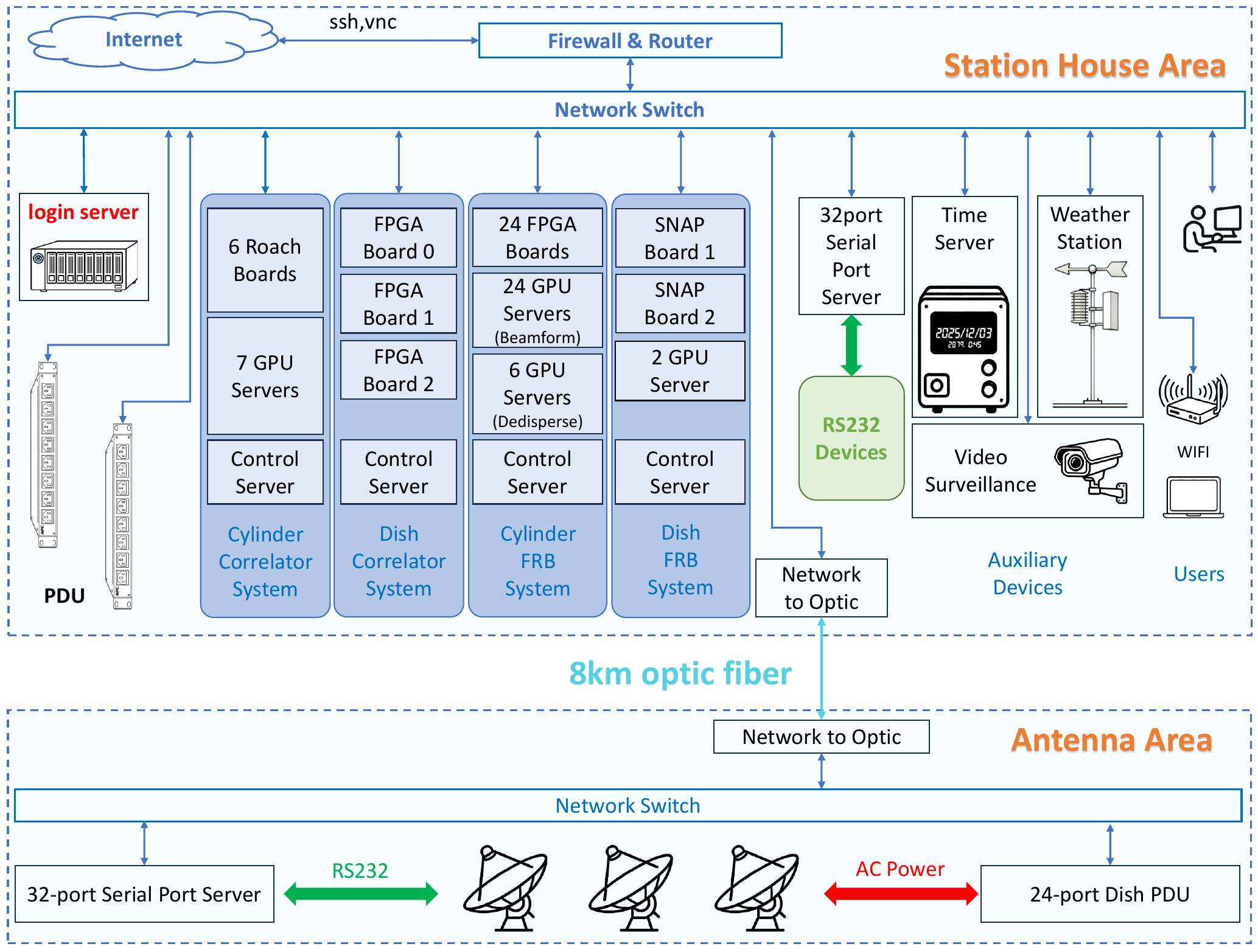}
    \caption{A schematic of the local area network and the various components. }
    \label{fig:lan_schematic}
\end{figure*}

A local area network (LAN) is set up to connect all the devices in the station house and antenna site, and it served as the backbone for our control system. A schematic of the LAN is shown in Fig. \ref{fig:lan_schematic}. 

A router located at the station house is used to set up the LAN and connect it to the internet. It has an embedded firewall service to protect the network. A dedicated server, which is called {\it login server} below, is setup for unified access to the site network and management of many devices and servers. It runs a Linux operating system, and provides ssh (secure shell) and vnc (virtual network console) services, so that it can be accessed from both within the LAN and from the outside through the Internet. Many of the control softwares for the auxiliary devices, mostly written in Python or shell scripts, are deployed in this login server. Security techniques such as password-protection, fail2ban, etc. are employed to protect this server against malicious attacks.

For the 4 sets of digital systems, as there are a large amount of internal data exchange, they are formed as private sub-net using network switches. 
The corresponding control server works as the gateway to connect the sub-net to the site LAN. To manage the observing system, an observer must first log in the login server, and then from there log in the corresponding control server. This physical isolation enhance the security level of the whole system, and also reduces the network traffic for the site LAN.

Within the LAN, the servers and devices are assigned with static IP addresses. 
The whole IP network is 192.168.1.0/24. The Router takes the IP address of 192.168.1.1 and acts as the DHCP server. Static IP addresses are assigned for various servers and their corresponding baseboard management controller (BMC) for remote access via web pages, various remotely-controlled devices at both the station house and antenna site. 
Users can also connect to the LAN using WiFi networks, which are sub-networks and does not take up the static IP address of the LAN.

Some devices are managed remotely using the RS-232 protocol and RS-485 protocol. For these devices, we have setup a 32-port serial port server (32p SPS) to bridge between the RS-232 and the LAN (for the dish antennas, a separate 32p SPS is setup), and a transfer device to  connect the RS485 devices to the 32p SPS.

\section{Power Control}
\label{sec:power}

\subsection{Power Distribution Units}

\begin{figure*}
    \centering
    \includegraphics[width=0.8\linewidth]{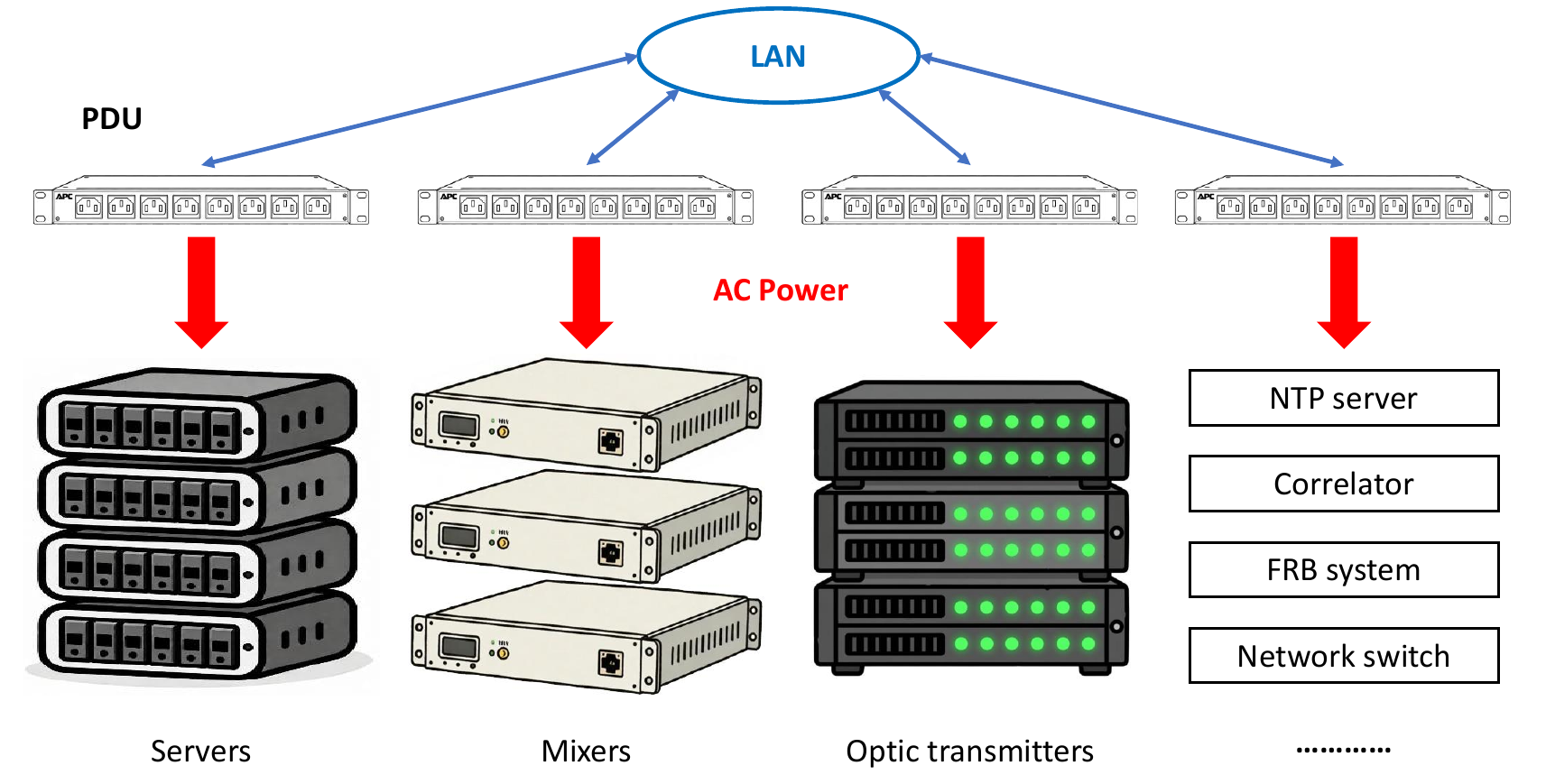}
    \caption{Various devices and servers are connected to PDUs which are remotely manageable through the LAN.}
    \label{fig:pdu}
\end{figure*}

All of the AC power of the servers and devices are designed to be able to be turned on and off remotely. The devices are configured to start working automatically once they are powered on, so the ON/OFF  control of each device can be realized by a remotely controlled power distribution unit (PDU) for its alternating current (AC) power. Commands can be issued via the LAN, which connects all PDUs, as shown in Fig. \ref{fig:pdu}.

The APC AP7921 made by Schneider Electric is selected as the PDU. This is widely used in the electrical industry for its reliability. It has eight C13 power sockets, each supports a maximum of 10 ampere current. It has a standard 1 unit enclosure for the 19 inch chassis, and can be installed conveniently. 
 
The online management of this PDU can be either through a password-protected graphical user interface (GUI), supplied  by the manufacture as a light-weight web server running inside the PDU, or through commands in the simple network management protocol (SNMP). However, the GUI interface has a relatively slow response and allows only one user to access it at one time, which is inconvenient, so in our case the SNMP is used for its control. 

The SNMP protocol is a standard application-layer protocol, it enables network administrators to remotely monitor device status, modify configurations, and receive fault alerts from devices of different vendors. The SNMP can be accessed by command line and supports muti-user access. Using the Python module {\tt pysnmp}, we have developed the program \texttt{pdu\_snmp.py} to manage the PDUs. This program is also deployed in the login server. It can be used to turn on/off and reboot specified outlets. It can also read the total electric current of the PDU in order to monitor potential overload event.

\subsection{Uninterruptible Power Supply (UPS)}
\label{subsec:UPS}

The UPS is a commonly used equipment to provide temporary power when the AC power is out by switching to batteries, and it is useful in filtering out voltage spikes, adjusting over-voltage or voltage sag to standard values, correcting instability of the mains frequency, correcting harmonic distortion and so on. We have installed a 100 kVA UPS for the digit department and analog department. A group of 44 100Ah batteries provides short time power when there is an power outage. However, if the power outage lasts more than 15 minutes, the continuous discharging of the batteries may cause damage by overheating the batteries. An automatic shutdown mechanism is needed. 
The UPS has a serial port interface which supports the RS-232 protocol, and is connected to the 32-port SPS. We developed the program \texttt{UPS.py} to communicate to the UPS through the SPS and deployed it in the login server. The commands are defined by the manufacture. Using this program, we are able to read out the running information of the UPS, such as the AC three-phase voltages and currents, the battery backup time, and the total power, etc. 

Additionally, a running program \texttt{emergency\_stop.py} calls \texttt{UPS.py} every 10 seconds to read the input voltage of the UPS, and check if one of the three-phase voltages is too low, e.g. below 200 V. Once this occurs, the program starts to record the total elapsed time of the abnormal voltage. Once this is over 15 minutes, it begins to execute emergency shutdown procedure. For devices, the program will call \texttt{pdu\_snmp.py} to power them off. For servers which are running operating systems, it will log into the servers and execute shutdown. Finally, the program shut down the login server where the \texttt{emergency\_stop.py} program runs in. An automatic emergency stop is thus accomplished. During the whole shutting down process, the program will take down the execution result into a log file and also attempt to send out an alert email to the designated administrator.

\subsection{Power in the Antenna Site}

In the antenna site, the devices which need power include the low noise amplifiers (LNA) and RF optical fiber transmitters, the calibration noise-source (CNS), and the Tianlai dish antenna drive motors and associated network devices for their sensing and control. The LNA, optical transmitter and the CNS need constant power during observations, and they are well designed to not generate RFI. The dish drive motors generate some RFI when it is powered on, though we have recently made some upgrade to reduce it. As noted early, the dish antennas are steerable, but we mostly do drift scan observations, during which the dish antennas are pointed to a specific direction, without the need of dish drive motors, so these can be powered off during the drift scan to eliminate their RFI. 

However, to realize this function, we also need to be able to control the power relays in the antenna site remotely, and in doing so, we need to avoid introducing new devices in the antenna site which generate RFI. In particular, many digital devices can produce RFI and should be avoided.

\begin{figure}
    \centering
    \includegraphics[width=0.8\linewidth]{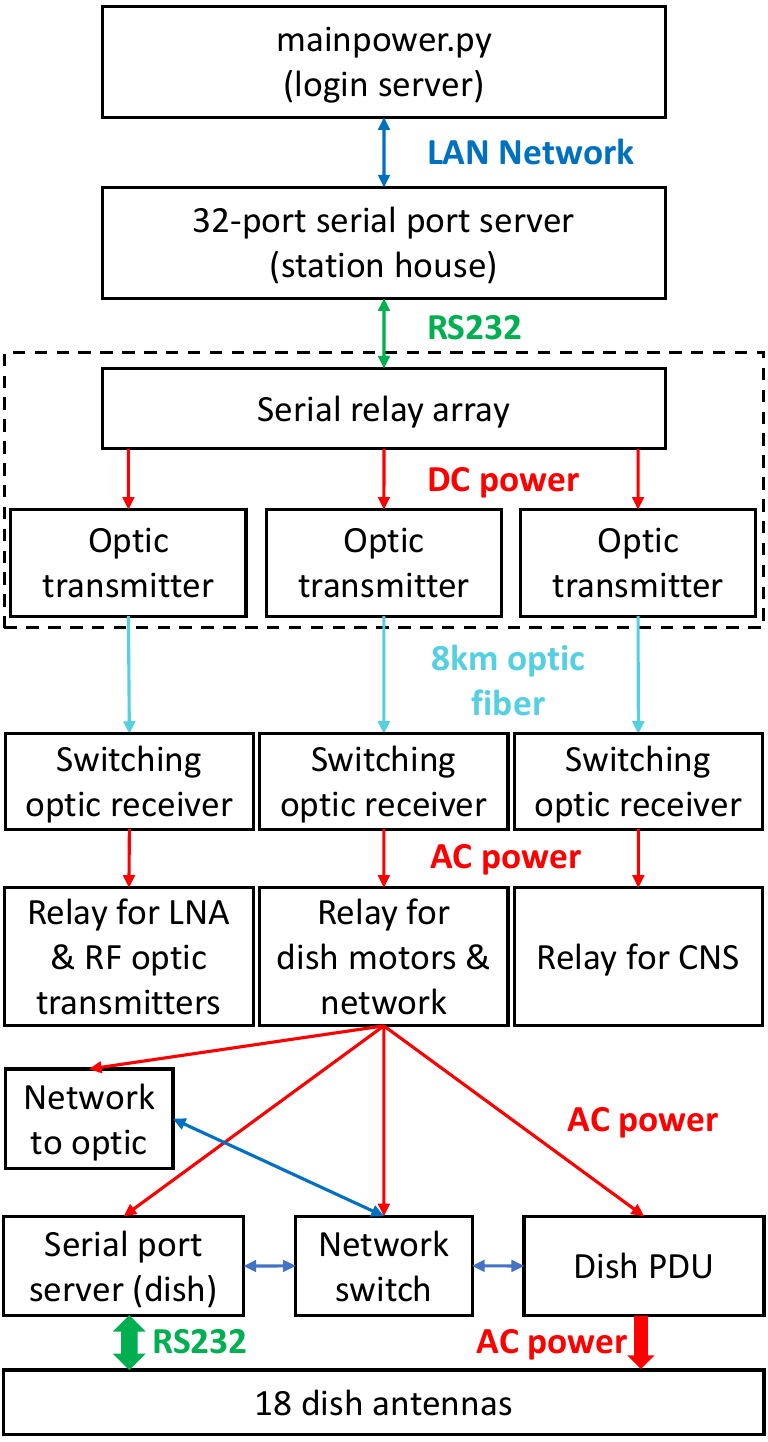}
    \caption{Main power control schematic in antenna site.}
    \label{fig:mainpower}
\end{figure}

Based on these considerations, we designed a power control system for devices in the antenna site as shown in Fig. \ref{fig:mainpower}. 
The dish motors and related network devices for sensing and controlling of the steering are connected to one power relay. The LNA and RF optic transmitters are connected to another power relay. The separately located CNS has its own power relay. Each relay is controlled by a corresponding switching optic receiver, which detects if there is optic signal or not in the fiber. These switching optic receivers and the relays have no digital electronics, thus RFI is avoided. Their power on/off are controlled from the Station House by sending optical signals. 

The powers are controlled by a serial relay array at the Station House. The serial relay array supports the RS-232 protocol, and is also controlled by the 32-port serial port server. In the login server, we developed the program \texttt{mainpower.py} to send the ON/OFF command through the LAN to the 32p SPS. Once the ON command is sent from the login server using TCP protocol, it is transferred to the serial relay array, which will power the optic transmitter. Then an optical signal is sent via the 8 km optic fiber, and it will be detected by the switching optic receiver. The receiver then latches the corresponding power relay. Our simple test show that the response time is about 0.15 second.

Once the relay for the dish motors and network devices is powered on, the network to optic converter, the network switch, a 32 port serial port server (dish-SPS) at the antenna site, and a dish antenna PDU begin to boot up. 
The network-to-optic converter are used to connect the devices in the antenna site to the LAN in the Station House so that every device is remotely manageable, and data on the dish antenna movement can be obtained. The dish-SPS is used to communicate with dish antennas using the RS-232 protocol and transfer the data into network packages. The dish PDU is a self-made device used to provide AC power for all dish antennas. Its main part is a 24-port relay array which supports network remote management. Compared to APC PDU, its main feature is a very fast boot speed of about 7 seconds, much shorter than the APC PDU boot time of 50 seconds, though it does not have the GUI interface of the APC PDU. The total boot up time is determined by the longest one from the network switch, which is about 9 seconds. 
In the login server, we developed the program \texttt{pdu\_self.py} to talk with the 24-port relay array using UDP protocol. The program supports checking and switching on/off for the power of each antenna.

\section{Dish Antenna Control}
\label{sec:ant_ctrl}

The steering of the dish antenna is controlled by the commands defined by the manufacturer through the RS-232 protocol. We use the dish-SPS to connect to all dishes in the antenna site. As the RS-232 protocol only supports a communication distance of about 30 meters, the serial server is installed in the center of dish antenna array. 

\begin{figure*}
    \centering
    \includegraphics[width=0.8\linewidth]{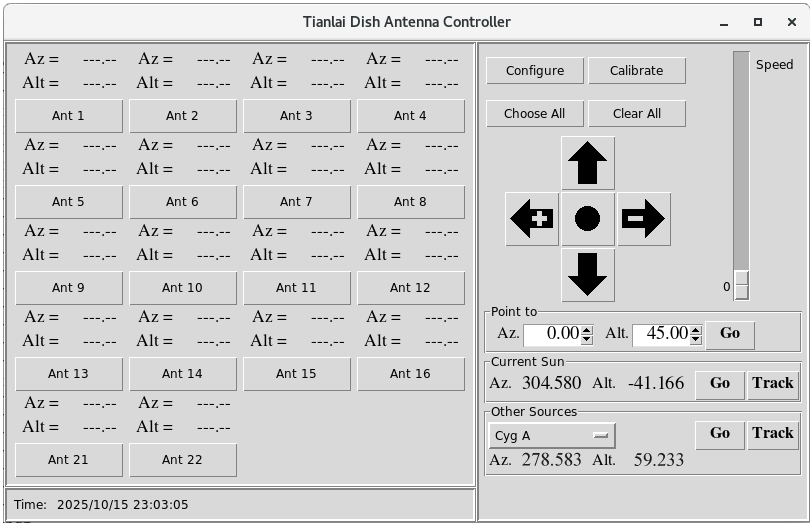}
    \caption{GUI of the dish antenna controller}
    \label{fig:dishpanel}
\end{figure*}

We developed the code \texttt{dishpanel.py} in the login server to remotely control the steering of the dish antennas, by communicating through the dish-SPS. \texttt{dishpanel.py} has a graphic user interface (GUI), which displays the pointing azimuth and altitude for each antenna, as shown in Fig. \ref{fig:dishpanel}. Commands can be issued to point a given antenna to a specific direction, either by using this GUI interface, or by using a simple command line program \texttt{antenna\_send.py}. The server then slews the antenna towards the target position, and returns to the control once the position is reached. There are also commands which allow multiple antennas to move toward the same position.

The pointing of the antenna is sensed by using an encoder on the drive motor, so it needs to be calibrated first. The program \texttt{dishpanel.py} has the function of directional calibration. The dish antenna is of 6 meter in diameter. In current frequency band 700--800~MHz, the beam width is about 4 degrees \citep{Wu2021dish,Zhangjy2021dishbeam}, so the requirement for the directional calibration precision is relatively low. We have adopted a simple method to make this directional calibration. The feed is installed at the focus point of the dish, which is along its central axis. When the shadow of the feed under the Sun is projected onto the center of the dish, we can assume the dish is pointing toward the Sun. At this moment, one can click the \texttt{Calibrate} button, the program will correct the goniometer to the right values as follows:
\begin{align*}
    \alpha_\mathrm{new} &= \alpha_\mathrm{sun} - \alpha_\mathrm{ant} + \alpha_\mathrm{old} \\
    \phi_\mathrm{new} &= \phi_\mathrm{ant} - \phi_\mathrm{sun} + \phi_\mathrm{old}\\
\end{align*}
where $\alpha_\mathrm{old}$ is the old azimuth correction value, $\alpha_\mathrm{sun}$ is the azimuth of the Sun and $\alpha_\mathrm{new}$ is the new azimuth correction value. The altitude $\phi$ is similar to the azimuth $\alpha$. Once calibrated, the correction values are automatically updated to the configuration file. Test radio observations of celestial sources show that the calibration is sufficiently accurate for our purpose.  

The dishes can also track celestial radio sources, although the tracking precision is not very high, and in the tracking mode, the motors may generate some RFI. The \texttt{dishpanel.py} program also provides a tracking mode. In this mode, one can choose a frequently used radio source such as Cyg A, Cas A, Virgo A, etc., the antenna control program will automatically compute the corresponding azimuth and altitude of the tracked source, and make updates with a period of 5 seconds, and move the antenna to the new direction.

\section{Array Operation Control}
\label{sec:array}

In this section, we introduce the operation control for the analog and digit department in the station house. For both the cylinder and dish arrays, there are electronically controlled attenuators for each signal channel, so as to equalize the signal strength of each channel; and there are also the local oscillators, which are adjusted to set the observing frequency band. The control programs of these are deployed in the login server. Each of the cylinder and dish arrays also has both a correlator to generate interferometric visibilities, and an FRB backend to search for FRBs.  
Due to different hardware architectures, each of these 4 digital backend has its own control server. Below, each of these programs are presented.

\subsection{Channel Equalization}

While all signal chains have the identical design and construction, the signal level for each receiver still differs from each other. The Tianlai signal chain has a large gain of about 100 dB, so at the end the difference can be significant. 
Electronically-controlled attenuators are used for equalization. Each group of 12 receiver channels is enclosed together into one chassis, and a total of 16+1 (with one redundant backup) chassis are deployed. 
To manage these attenuators, we developed a program (\texttt{receiver.py}). These attenuators are connected to the 32p SPS using the RS232 protocol, and the SPS waits for UDP connections on the network. The different attenuators are distinguished by their network ports. In this way, the program \texttt{receiver.py} can simply connect to the SPS using UDP protocol, and communicate with each receiver by sending commands to different UDP ports. 

The program \texttt{receiver.py} is an object oriented program. Two classes are defined. First, the class \texttt{Receiver} corresponds to one single receiver with support of checking and setting attenuation for all 12 channels, and saving and loading settings to and from file. Second, the class \texttt{AllReceiver} is combination of all 17 receivers based on \texttt{Receiver}.

The program has the function of checking and setting  the attenuation of each channel. The basic procedure to adjust the attenuation for all signal channels is as follow. First, we choose a time at night when no bright source such as Cyg A is transiting the beam. Then, we measure the IF signal level of one channel (for convenience, usually channel 1 is chosen) using a spectrum analyzer. By adjusting the electronic attenuation of channel 1, the signal level of this channel is adjusted to a suitable value as discussed in Sec. 3.2 of Ref.\cite{Lijx2020tianlai}, usually -13 dBmW. Next, we do a trial observation and obtain the auto-correlation data of all channels. By comparing the auto-correlation of other channels to channel 1, the attenuation required for each channel are obtained. Concretely, if we denote the frequency averaged auto-correlation of channel $n$ by $I_n$, then the required extra attenuation is $\Delta A = 10 \times \log_{10}(\frac{I_n}{I_1})$. 
This procedure is usually repeated two or three times to confirm that all signal levels are equalized.

\begin{figure}
    \centering
    \includegraphics[width=0.78\linewidth]{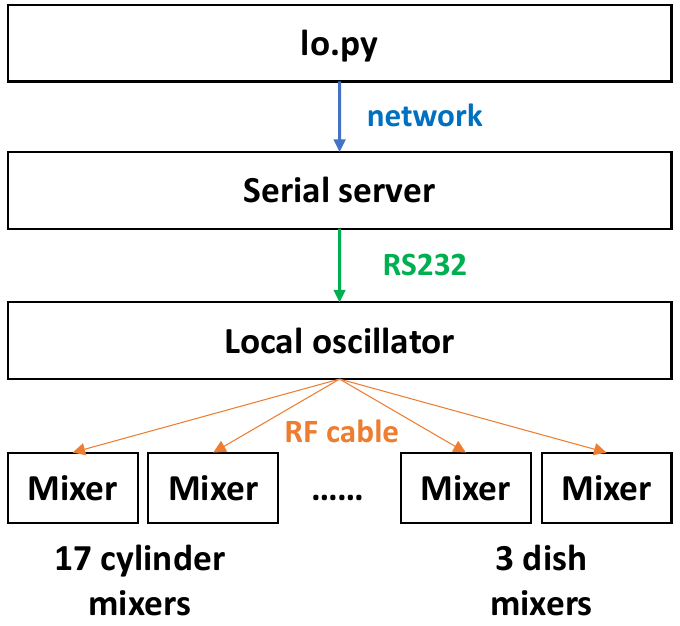}
    \caption{Local oscillator program logic.}
    \label{fig:lologic}
\end{figure}

\subsection{Local Oscillator}
\label{subsec:lo}

The local oscillator (LO) is an important device in the analog receiver system. It provides a stable frequency basis for all the 17 cylinder mixers and 3 dish mixers. 
The program control logic is shown in Fig. \ref{fig:lologic}. 
The management of LO is accomplished through the DB9 socket using the RS-232 protocol. This socket is also connected to the 32p SPS. 
We developed the program \texttt{lo.py} to adjust the frequency and signal output level of the LO. This program runs on the login server, which connects to the serial port server through LAN, and then the serial server forwards the command to the LO. Using \texttt{lo.py}, we can check current setting, set the LO frequency and attenuation.

\subsection{Programmed Observation and Calibration of the Dish Array}

The observation of the dish array involves pointing the antennas to specific directions according to an observation plan. This can be realized by automatic control program. 
In fact, even for a drift scan observation, the complex gains of the dish array need to be calibrated at some time intervals. The absolute calibration is realized by observing a bright celestial source, such as Cyg A or Cas A, which can be treated as a point source for the resolution of our array. So the dish antennas needs to be pointed to the calibration source on the sky, making the observation, then point back to the usual drift scan position.

\begin{figure}
    \centering
    \includegraphics[width=0.95\linewidth]{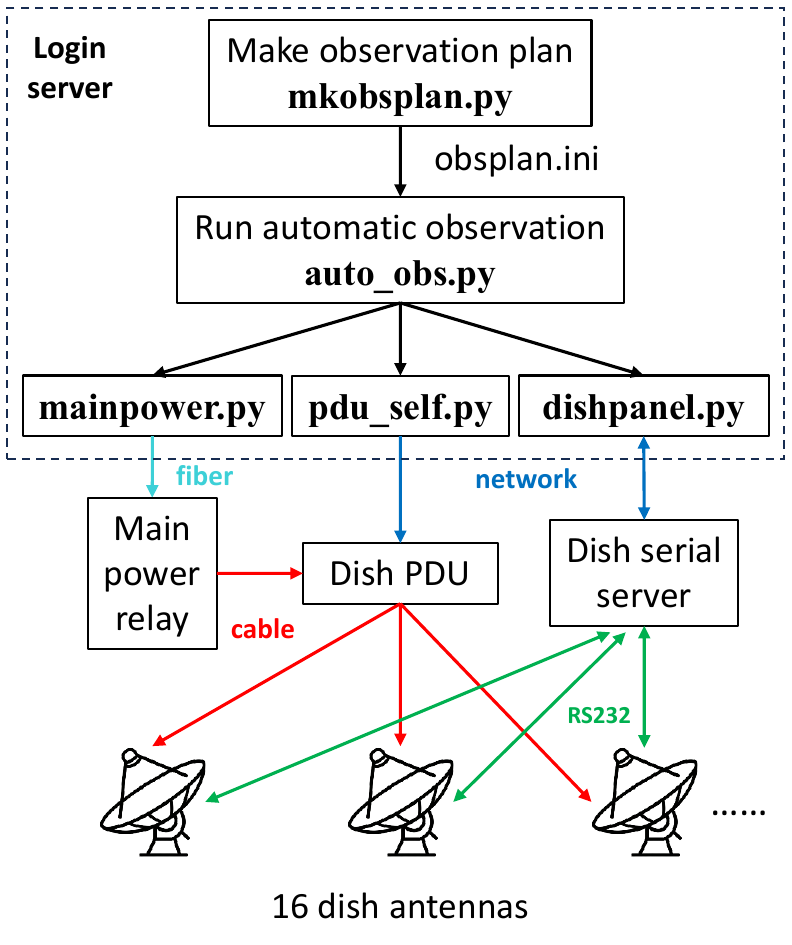}
    \caption{The program logic of the automatic observation.}
    \label{fig:autoobs}
\end{figure}

To carry out this procedure we developed an automatic observation program \texttt{auto\_obs.py}. As shown in Fig. \ref{fig:autoobs}, this program loads in the \texttt{obsplan.ini} file, which gives the observation plan -- when to point to which direction to observe which source. The observation plan can be generated by another program \texttt{mkobsplan.py}, which can calculate the azimuth and altitude for the frequently-used radio sources as seen at the antenna site at the planned observation time. The program executes the calibration observation, and 
once the calibration procedure is accomplished, the antennas are again pointed to the position for the drift scan. 
During this process, the program  \texttt{auto\_obs.py} invokes the main power management program \texttt{mainpower.py} and the power management program \texttt{pdu\_self.py} to power the dish drive mechanism, and 
the dish antenna control program \texttt{dishpanel.py} to point the dishes to the correct position, and turn off the power when doing the calibration or drift scan observations.

\subsection{Correlator Backend Control}

The Tianlai cylinder correlator backend is described in \citet{Wangzh2024roach}. The F-engine performs FFT and sends the transformed data to the X-engine where the data from different feeds but with same frequency are sent to the same GPU server for correlation and integration. The integrated correlation data are re-arranged by the control server, and dumped onto hard disks in the \texttt{HDF5} format as defined in \citet{Zuosf2021tlpipe}. 

The whole cylinder correlator system has been designed to be fully remotely controllable. The AC power is controlled by PDUs, and can be turned on or off remotely. Once the F-engine Roach boards are powered up and ready, the shell script \texttt{fengine.sh} is used to download the firmware to the FPGA chips to start their work. Next, the shell script \texttt{allpaper\_init.sh} is used to initialize all the GPU servers of the X-engine. Finally, by running the script \texttt{allstart.sh}, the GPU servers are synchronized to the same integration time and start to output the correlation data to the control server. In the control server, a program \texttt{chk\_port.py} can be run firstly to check if all the GPU server are working properly. If everything is Okay, one can finally starts the observation by running the script \texttt{roach\_vis\_192ch.py}. 

\begin{figure}
    \centering
    \includegraphics[width=0.8\linewidth]{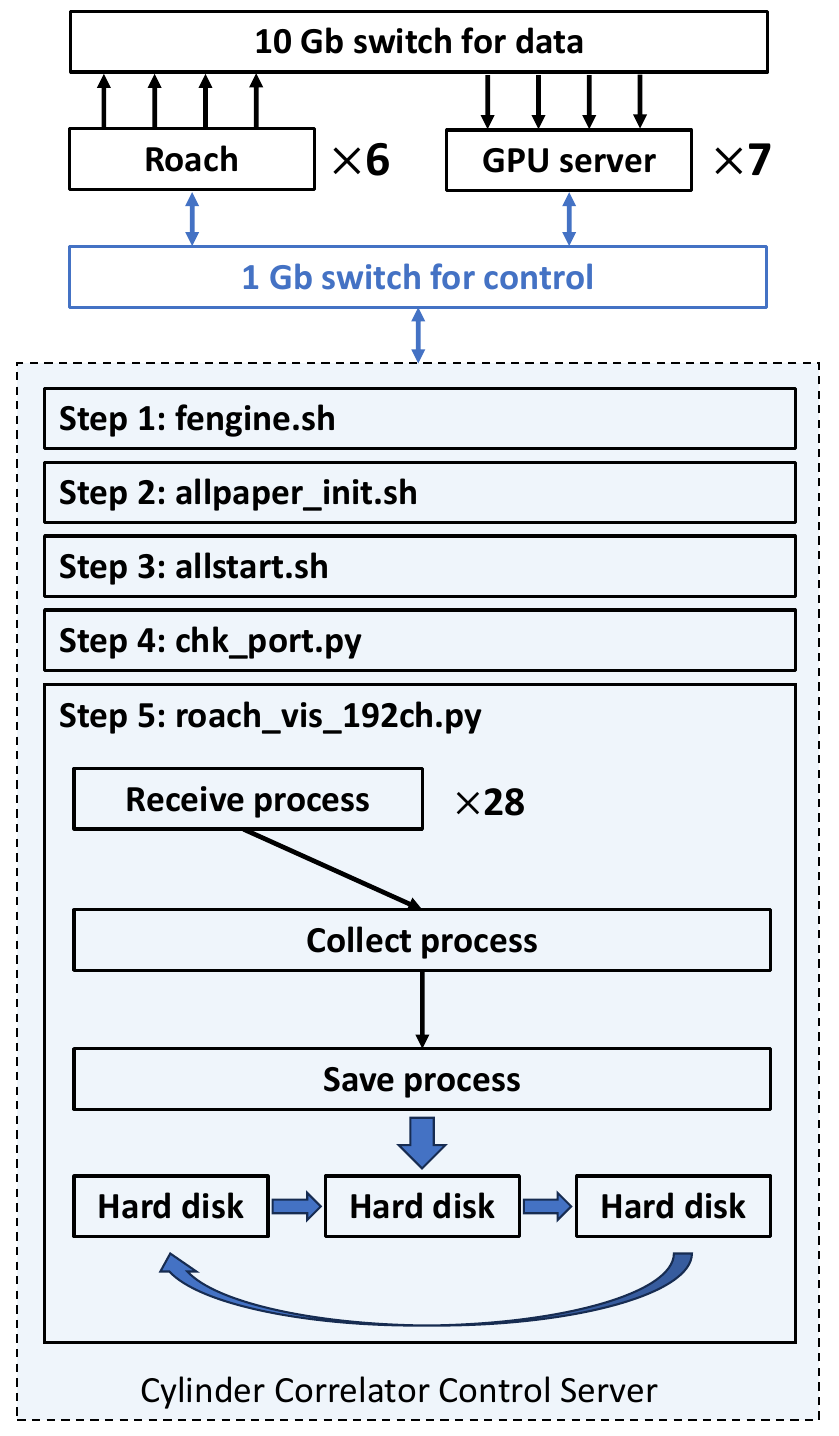}
    \caption{Program logic for cylinder correlator.}
    \label{fig:roach_vis}
\end{figure}

Each GPU server processes the correlation data for four fixed frequency channels. The data of the same frequency channel are sent to the same UDP port in the control server. 
In the program \texttt{roach\_vis\_192ch.py}, as shown in Fig. \ref{fig:roach_vis}, there are 28 \texttt{Receive} processes created. Each process is bound to one UDP port for receiving packets of the same frequency segment. Then data are transmitted to the \texttt{Collect} process. 
This process receives all the data from all the \texttt{Receive} process, and rearranges them according to their correct frequency sequence, and then sends the data to \texttt{Save} process. Finally, the \texttt{Save} process dumps the data into a group of hard disks in HDF5 format. As the data rate of the cylinder array is massive, the data are written into hard disks one by one. The \texttt{Save} process will use these hard disks in a cyclic way. Observers can replace those fully written disks with empty ones. If only one empty disk is left, the process will send out an email to notify the observers to replace the disks as soon as possible. This design ensures a long-term observation without interruption. Besides the visibility data, meta parameters such as the frequency value, site information, CNS information, etc. are also saved in the attributes of the HDF5.

\begin{figure}
    \centering
    \includegraphics[width=0.9\linewidth]{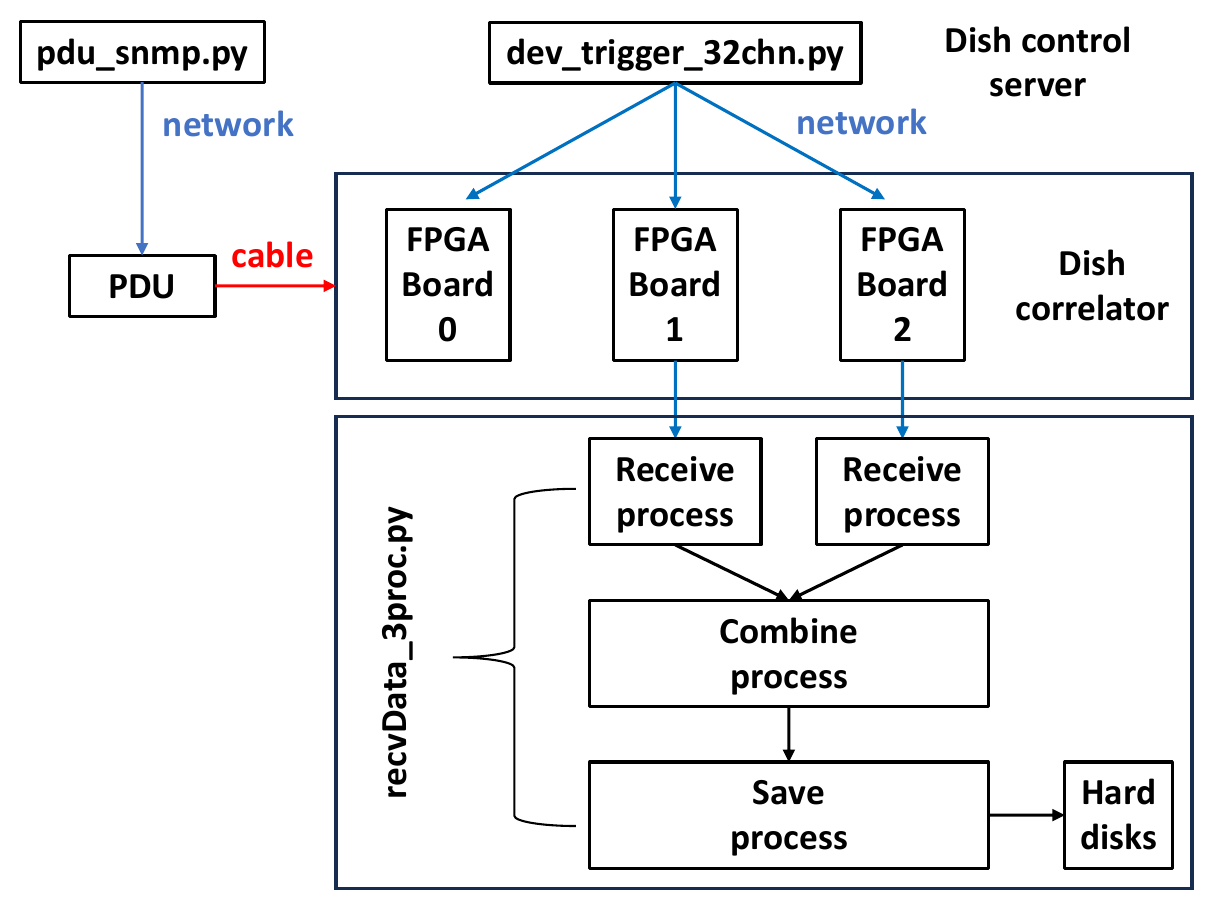}
    \caption{Dish correlator program logic.}
    \label{fig:dish_corr}
\end{figure}

The Tianlai dish correlator backend is controlled by the program \texttt{dev\_trigger\_32chn.py}, which is run on the control server of the dish correlator as shown in Fig. \ref{fig:dish_corr}. It sets up the running parameters such as the integration time, triggers the correlator to start working, and receives data from the correlator.  

The CNS is controlled by either of the correlator. To align the on/off period of the CNS to the integration time, a  Transistor-Transistor Logic (TTL) signal output from the correlator is used to control the on/off status of the CNS. For example, the program \texttt{dev\_trigger\_32chn.py} can set the on time, off time and period of the CNS. Once the trigger signal is sent, the correlator will enter into the working status and continuously output the correlation data. 
The observation data from the correlator are output through 2 Ethernet sockets. We developed the program \texttt{recvData\_3proc.py} to read and save the data into \texttt{HDF5} format. 
It creates 2 \texttt{Receive} processes to receive the packet data from the correlator. Data are combined and rearranged in the \texttt{Combine} process. Finally, the \texttt{Save} process will save the visibility data into HDF5 file on hard disks. 
Offline analysis of the data can be done using the Tianlai pipeline \texttt{tlpipe} \citep{Zuosf2021tlpipe}. 
Both the \texttt{dev\_trigger\_32chn.py} and \texttt{recvData\_3proc.py} use the Ethernet II frame protocol as defined by the correlator manufacture.

\subsection{FRB Backend Control}
\label{subsec:frb_backend}

The FRB backends are running simultaneously with the correlator backends. For the cylinder FRB backend, the control programs are deployed on the control server. 
The F-engine and the B-engine are controlled by a web based GUI management system which consists of three management webpages. 
In the F-engine management web page, the program provides the function to power ON/OFF the F engine, download the firmware, start the system and show the running status. It is also capable to get some sample data from the ADC or the F-engine for diagnosis. In the B-engine management web page, the program provides the function to power ON/OFF all GPU servers and start the beam forming program. In the resource monitoring web page, the program is able to monitor all the CPU, memory and GPU usage status. Once the F engine and B engine are up and running, beam forming data streams are then sent to the D engine for dedispersion and FRB search. The D engine is controlled by several command line programs. The program \texttt{basic\_initial.sh} is used to prepare the temporary data directories and mount the program directories for all GPU servers. Then, the program \texttt{hashpipe\_initial.sh} is used, which calls the program \texttt{TL\_init.sh} to create hashpipe instances to receive and process the data stream from the B engine. Next, the program \texttt{dedisperse\_init.sh} is started to dedisperse the beam forming data for a certain time interval (called a frame), and search for FRB candidate in each frame. If no candidate is found in a frame, the data are deleted. If one or more candidates are detected, that frame of data is then stored for further analysis.

For the dish FRB backend, the control mechanism is similar, though the power up of the F engine boards and the GPU servers are managed by the command line program \texttt{pdu\_snmpy.py}. After the boards and servers are ready, a script \texttt{dishfrb\_initial.sh} is used to initialize the F engine boards, setting parameters such as the integration time, the beam weighting coefficients, etc. Next, the program \texttt{sys\_trig.py} is used to trigger the F engine to start, and the delay time of each antenna is set using \texttt{adc\_adjust.py}. Then, hashpipe in D engine is started by running the script \texttt{Pipe\_run.sh}. This script will start hashpipe instances to dedisperse the beam forming data, search FRB candidate, and finally delete the data if no candidate is found, or save the data in long-term storage if any FRB candidate is detected.

\subsection{Auxiliary Data}
\label{sec:other}

\begin{table*}
    \centering
    \begin{tabular}{c|l|l|l}
    \hline\hline
    Facility & Name & Protocol & Function \\
    \hline
    Power & pdu\_snmpy.py & SNMP & PDU control program for various devices. \\
        &   pdu\_self.py & UDP & PDU control program for dish antennas.\\
        &   mainpower.py & TCP, RS232 & Main power control in antenna site.\\
         \\
     UPS   &   UPS.py & TCP, RS232 &  UPS control program.\\
        &  emergency\_stop.py & --- & Emergency stop program.\\
        \\
    Auxiliary &  tlweather.py & TCP, RS232 & Weather station control program.\\
        & record\_weather.py & --- & Weather data recording data program.\\
        & air\_conditioner.py & TCP, RS485 & Air conditioner control program.\\
        \\
    Dish Antenna & dishpanel.py & UDP, RS232 & Dish antenna control program.\\
    \\
    Digital Backends & recvData\_3proc.py & Ethernet II & Dish correlator data recording program.\\
        & fengine.sh & --- & Roach firmware download program.\\
        & allpaper\_init.sh & --- & Initialize the GPU servers of the X-engine \\
        & allstart.sh & --- & Cylinder correlator trigger program.\\
        & chk\_port.py & UDP & Check cylinder correlator ready program.\\
        & roach\_vis\_192ch.py & UDP & Cylinder correlator data recording program.\\
        & basic\_initial.sh & --- & Cylinder FRB system basic initialize program. \\
        & hashpipe\_initial.sh & --- & Cylinder FRB hashpipe initialize program.\\
        & TL\_init.sh & --- & Cylinder FRB hashpipe program. \\
        & dedisperse\_init.sh & --- & Cylinder FRB dedispersion program.\\
        & dishfrb\_initial.sh & --- & Dish FRB initialize program. \\
        & sys\_trig.py & --- & Dish FRB system trigger program. \\
        & adc\_adjust.py & --- & Dish FRB system delay setting program. \\
        & Pipe\_run.sh & --- & Dish FRB system dedispersion and search program. \\
        \\
    Array Operation & lo.py & TCP, RS232 & LO control program.\\
        & receiver.py & UDP, RS232 & Receiver control program.\\
        & mkobsplan.py & --- & Make plan for automatic observation.\\
        & auto\_obs.py & --- & Automatic observation program.\\
        & cns\_switch.py & TCP, RS232 & CNS control program.\\
        & dev\_trigger\_32chn.py & Ethernet II & Dish correlator trigger program.\\
    \hline\hline
    \end{tabular}
    \caption{A list of the control programs and their functions.}
    \label{tab:program_list}
\end{table*}

We have also recorded some auxiliary data. For example, the system gain can be affected by the ambient temperature, with lower gain for higher temperature. 
To record the temperature of the analog department room, we directly read the data from the A/C. A Joton industrial A/C is used to cool the analog department room. The program \texttt{air\_conditioner.py} communicates with the A/C, to read out the temperature and humidity of the room, and other data such as the power on/off, power status, working hours of the air conditioner. This communication is also accomplished via the 32p SPS. 

To obtain the outdoor ambient temperature, we have installed a weather station outside the Station House. The program \texttt{tlweather.py} is used to communicate with the weather station, which reads out various weather data such as the temperature, humidity, dew point, wind speed and so on. The weather station has an embedded serial server which transfers RS-232 data into TCP data. The program \texttt{tlweather.py} can thus connect to the TCP server and obtain the weather data.

The program \texttt{record\_weather.py} is used to record the temperature data into files. It periodically recalls \texttt{air\_conditioner.py} and \texttt{tlweather.py} to obtain various temperature data and saves to text files. A new file is generated everyday.

\section{Conclusion}
\label{sec:conclusion}

This paper presents the design, implementation and on-site deployment of the complete operation control system for the Tianlai 21cm intensity mapping experiment. The system directly solves two coupled core challenges at the remote Hongliuxia Observing Station: (1) reducing the high manpower and travel costs incurred by the geographical isolation of the site; and (2) the critical need to protect the quiet electromagnetic environment to prevent additional RFI from control equipments from overwhelming the extremely weak cosmic 21cm neutral hydrogen signals. 

To address these challenges, we developed an integrated hardware–software control system based on five foundational principles: online accessibility, simplicity, heterogeneous device compatibility, strict RFI suppression, and network security. The system adopts a hierarchically isolated LAN architecture, accessible via the internet, which serves as the unified command and control backbone. Through this network, we have implemented a full-stack solution for the operation of the telescope and other facilities at the observing station. 
A key achievement of this work is the realization of full-process remote controllability over nearly all facilities and instruments at the Hongliuxia Station. As detailed in Table \ref{tab:program_list}, we have developed a suite of dedicated programs that enable remote operation of the Tianlai experiment. 
This capability translates into remarkable practical outcomes: the system supports end-to-end remote observations, with very low on-site manpower. The innovative optical power control scheme completely eliminates motor-induced RFI during observations, safeguarding the station’s pristine electromagnetic environment. Our system drastically reduces operational costs and safety risks associated with on-site maintenance in harsh winter conditions, establishing itself as a critical technical pillar for the long-term scientific success of the Tianlai experiment. 

For future improvements, we propose the following directions: (1) Establish a centralized and unified log platform, integrated with an intelligent fault detection and early warning algorithms; (2) Upgrade the alert system to a multi-channel, hierarchical framework (incorporating instant messaging, SMS, and phone calls) to shorten administrator response times; (3) Develop a closed-loop intelligent observation system that autonomously adjusts strategies based on real-time data quality feedback; and (4) Extend the system’s modular compatibility to newly deployed facilities, such as the global spectrum measurement system and FRB outrigger arrays, to form a universal and scalable control framework.

In summary, the lightweight, low-RFI, and highly compatible architecture validated in this work provides a reliable, replicable, and cost-effective reference for the control system design of other remote astronomical observatories and ground-based scientific facilities worldwide.

\section*{acknowledgements}
We thank the 21CMA, AIMS and MUSER teams, esp. Xianyong Bai, Linjie Chen, Jing Du, Junhua Gu, and Yan Huang for informing us on the status of their experiments.
This work is supported by National SKA Program of China (Nos. 2022SKA0110100 and 2022SKA0110101), the National Natural Science Foundation of China (No. 12203061, 12361141814, 12303004, 12273070, 12473094), and the Chinese Academy of Science ZDKYYQ20200008. 

\section*{Author Contributions}
Jixia Li designed and implemented the controlling system described in this paper, and prepared the draft. Fengquan Wu leads the design and construction of the Tianlai systems, Fengquan Wu, Shijie Sun and Yougang Wang deployed the devices and instruments. Xuelei Chen is the leader of the Tianlai project and made detailed revision of the paper. All authors read and approved the final manuscript. 

\section*{Software Availability}
The software described in this paper is available freely from the authors upon reasonable request.
 
\section*{Declaration of Interests}
Xuelei Chen is the editorial board member for Astronomical Techniques and Instruments. He was not involved in the editorial review or the decision to publish this article. The authors declare no competing interests.

\section*{AI disclosure statement}

\begin{figure}
    \centering
    \includegraphics[width=0.9\linewidth]{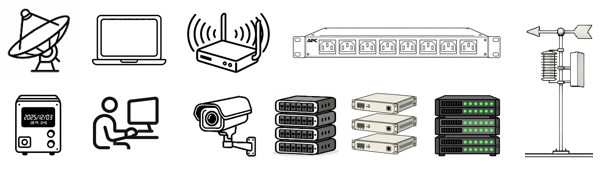}
    \caption{Logos drawn by AI.}
    \label{fig:ai_logos}
\end{figure}

AI-assisted technology is used in the preparation of this work. The usage is limited in drawing a few logos of the schematic graphs as shown in Fig. \ref{fig:ai_logos}. All other works are done without AI assistant. 

\appendix


\clearpage

\bibliographystyle{ati} 
\bibliography{ati}      

\label{lastpage}

\end{document}